\documentclass[epj]{svjour}

\usepackage{graphicx}
\usepackage{colordvi}
\usepackage{epsf}
\usepackage{amsmath}

\abstract{ The semileptonic decay $B\to\pi$ is studied starting from a
simple quark model that takes into account the effect of the $B^*$
resonance.  A novel, multiply subtracted, Omn\`es dispersion relation
has been implemented to extend the predictions of the quark model to
all $q^2$ values accessible in the physical decay. By comparison to the
experimental data, we extract
$|V_{ub}|=0.0034\pm0.0003{\rm(exp.)}\pm0.0007{\rm(theory)}$. As a further
test
of the model, we have also studied $D\to\pi$ and $D\to K$ decays for which we
get good agreement with experiment.
}

\title{Study of the semileptonic decays $B\to\pi$, $D\to\pi$ and $D\to
K$} \author{C. Albertus \inst{1}, J. M. Flynn \inst{1},
E. Hern\'andez \inst{2}, J. Nieves \inst{3}, J. M. Verde-Velasco
\inst{2} }

\institute{School of Physics and Astronomy, University of Southampton,
Southampton SO17 1BJ, United Kingdom. \and Grupo de F\'\i sica
Nuclear, Departamento de F\'\i sica Fundamental e IUFFyM, Facultad de
Ciencias, E-37008 Salamanca, Spain. \and Departamento de F\'\i sica
At\'omica, Molecular y Nuclear, Universidad de Granada, E-18071
Granada, Spain.}

\date{Received: date / Revised version: date}

\begin{document}

\authorrunning{C. Albertus {\it et al.}}

\maketitle

\section{Introduction}

The  exclusive semileptonic decay $B\to\pi l^+\nu_l$
provides an important alternative to 
 inclusive reactions  $B\to X_u
l^+\nu_l$ in the determination of de Cabibbo-Kobayashi-Maskawa (CKM) matrix
element $|V_{ub}|$. 

This reaction has been studied in different approaches like
lattice-QCD (both in the quenched and unquenched
approximations), light-cone sum rules (LCSR) and  constituent quark
models (CQM), each of them having 
 a limited range of applicability: LCSR are suitable
for describing the low  momentum transfer square ($q^2$) region, while
lattice-QCD provides results only in the high $q^2$ region. CQM can in principle
provide form factors in the whole $q^2$ range but they are not directly
connected to QCD.  A combination of different methods seems
to be the best strategy. 

The use of Watson's theorem for the $B\to\pi l^+ \nu_l$ process allows one to
write a dispersion relation for each of the form factors entering in
the hadronic matrix element. This procedure leads to the so-called
Omn\`es representation, which can be used to constrain the $q^2$
dependence of the form factors from the elastic $\pi B\to \pi B$
scattering amplitudes. The problem posed by the unknown 
 $\pi B\to \pi B$ scattering amplitudes at high
energies can be dealt with by using a multiply
subtracted dispersion  relation. The latter will allow for the combination 
of  predictions from various 
methods in
different $q^2$ regions.

In this work we study the semileptonic $B\to\pi l^+ \nu_l$ decay. The use of a
multiply subtracted Omn\`es representation of the form factors will allow us 
to  use the predictions of LCSR calculations at $q^2=0$ 
in order to extend the results of a simple nonrelativistic constituent 
quark model (NRCQM) from its
region of applicability, near the zero recoil point, to the whole 
physically accessible $q^2$ range.
To test our model we shall also study the $D\to\pi$ and $D\to K$ semileptonic 
decays for which
 the
relevant CKM matrix elements are well known and there is precise experimental 
data.

\section{$B\to\pi l^+\bar{\nu}$}

The matrix element for the semileptonic $B^0\to\pi^-l^+\nu_l$ decay
can be parametrized in terms of two dimensionless form factors
\begin{align}
\left<\pi(p_\pi)\left|V^\mu\right|B(p_B)\right>=&
\left(p_B+p_\pi-q\frac{m_B^2-m_\pi^2}{q^2}\right)^\mu f^+(q^2)\nonumber\\&
+q^\mu\frac{m_B^2-m_\pi^2}{q^2} f^0(q^2)
\label{eq:matele}
\end{align}
where $q^\mu=p_B-p_\pi$ is the four momentum transfer and $m_B=5279.4$
MeV and $m_\pi=139.57$ MeV are the $B^0$ and $\pi^-$ masses. For
massless leptons, the total decay width is given by
\begin{align}
\Gamma(B^0\!\!\to\!\pi^-l^+\nu_l)=\!
\frac{G_F^2|V_{ub}|^2}{192\pi^3m_B^3}\int_0^\infty
\!\!\!\!\!dq^2[\lambda(q^2)]^{\frac32}|f^+(q^2)|^2
\end{align}
with $q^2_{\rm max}=(m_B-m_\pi)^2$, $G_F=1.16637\times 10^{-5}$ GeV$^{-2}$
and $\lambda(q^2)=(m_B^2+m_\pi^2-q^2)^2-4m_B^2m_\pi^2=4m_B^2|\vec{p}_\pi|^2$,
with $\vec{p}_\pi$ the pion three-momentum in the $B$ rest frame.
\subsection{Nonrelativistic constituent quark model: Valence quark 
and $B^*$ resonance contributions}
\begin{figure}
\begin{center}
\includegraphics[clip,scale=0.6, bb= 250 500 520 800]{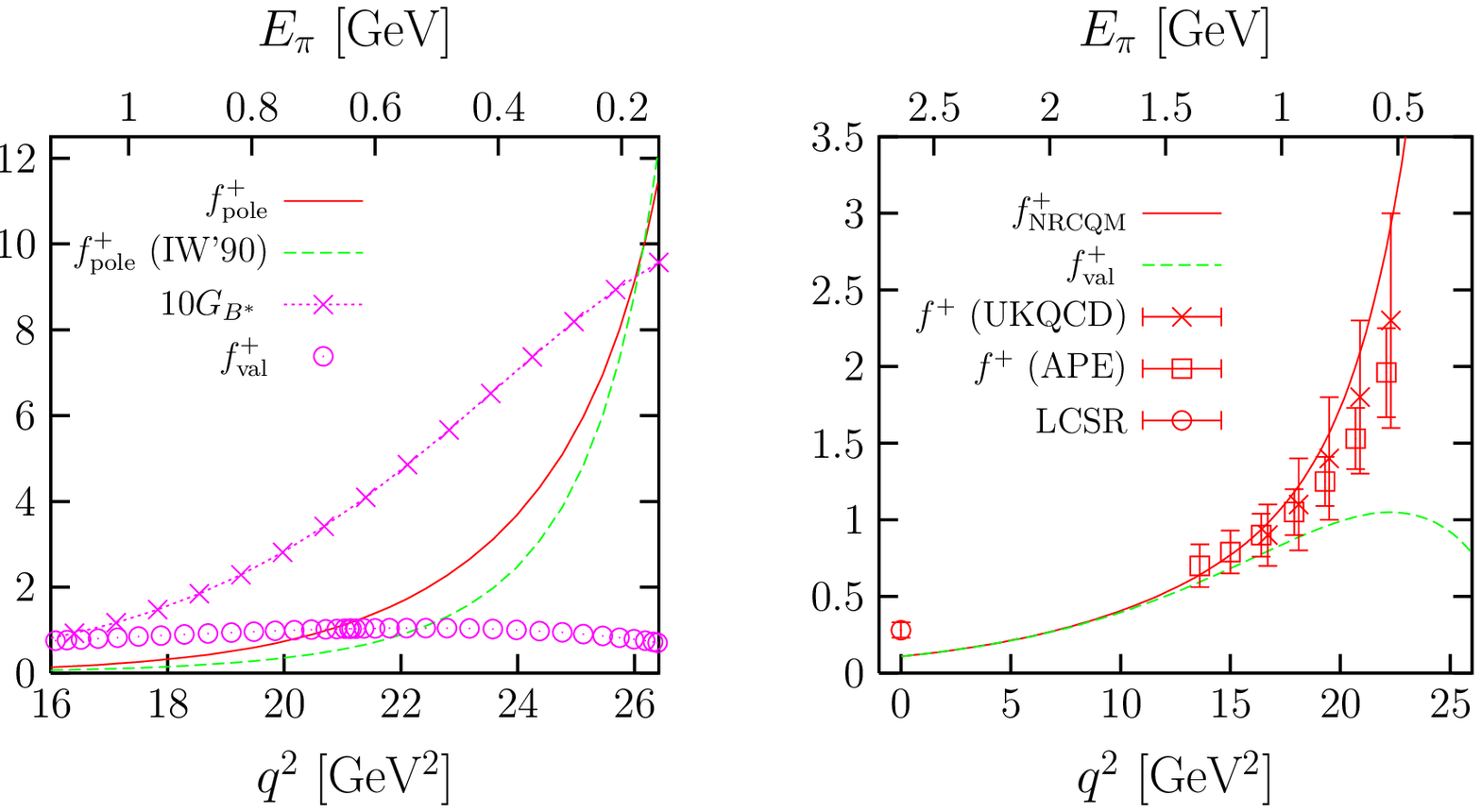}
\caption{$f^+$ form factor obtained with the valence quark (val)
contribution alone and with the valence quark plus $B^*$
contribution (NRCQM). We also plot lattice QCD results by the UKQCD~\cite{ukqcd} and
APE~\cite{ape} Collaborations, and
LCSR~\cite{lcsr} $f^+$
results.}
\label{fig1}
\end{center}
\end{figure}
%

Figure~\ref{fig1} shows how the naive NRCQM valence quark description of the
$f^+$ form factor fails in the whole $q^2$ range. In the region close to
$q^2_{\rm max}$, where a nonrelativistic model should work best, the
influence of the $B^*$ resonance pole is evident. Close to $q^2 =0$ the
pion is ultra relativistic, and thus predictions from a nonrelativistic
model are unreliable.

As first pointed out in Ref.~\cite{2}, the effects of the $B^*$ resonance pole
dominate the $B\to\pi l^+\nu_l$  decay near the zero recoil point ($q^2_{\rm
max}$). Those effects  must be added coherently as a distinct contribution 
to the valence result.
%
%
The hadronic amplitude from the $B^*$-pole contribution is given by
\begin{align}
&-iT^\mu=\\
&-i\hat{g}_{B^*B\pi}(q^2)p^\nu_\pi\left(i\frac{-g^\mu_\nu+q^\mu
 q_\nu/m_{B^*}^2}{q^2-m_{B^*}^2}\right)i\sqrt{q^2}\hat{f}_{B^*}(q^2)\nonumber
\end{align}
with $m_{B^*}=5325$ MeV. $\hat{f}_{B^*}$ and
$\hat{g}_{B^*B\pi}$ are respectively the off-shell $B^*$ decay constant 
and off-shell strong $B^*B\pi$
 coupling constant. See Ref.~\cite{1} and references therein for details on
 their calculation. From the above equation one can easily obtain the $B^*$-pole
 contribution to $f^+$ which is given by
\begin{equation}
f^+_{pole}(q^2)=\frac{1}{2}
\hat{g}_{B^*B\pi}(q^2) \frac{\sqrt{q^2}\hat{f}_{B^*}(q^2)}{m^2_{B^*}-q^2}
\end{equation}

The inclusion of the  $B^*$ resonance contribution to the form factor
 improves the simple valence quark prediction down
to $q^2$ values around $15$ GeV$^2$. Below that the description
is still poor.

\subsection{Omn\`es representation}

Now one can use the Omn\`es representation to combine the NRCQM
predictions at high $q^2$ with the LCSR at $q^2=0$. 
%
%
%
This representation requires as an input
the elastic $B\pi\to B\pi$ phase shift $\delta(s)$ in the $J^P=1^-$
and isospin $I=1/2$ channel, plus the form factor at different $q^2$ values 
below the $\pi B$ threshold where the 
 subtractions will be  performed.  For a large enough number of  subtractions, 
only the phase shift at or near threshold is needed. In that case
 one can approximate 
 $\delta(s)\approx\pi$, arriving at the result that
\begin{equation}
f^+(q^2)\approx\frac{1}{s_{th}-q^2}\prod_{j=0}^n[f^+(q^2_j)
(s_{th}-q^2_j)]^{\alpha_j(q^2)},\  n\gg 1
\label{eq:omnesrepapprox}
\end{equation}
with $s_{th}=m_B+m_\pi$ and 
$\alpha_j(q^2)=\prod_{j\ne k=0}\frac{q^2-q_k^2}{q_j^2-q_k^2}$
\begin{figure}
\begin{center}
\resizebox{0.75\columnwidth}{!}{\includegraphics{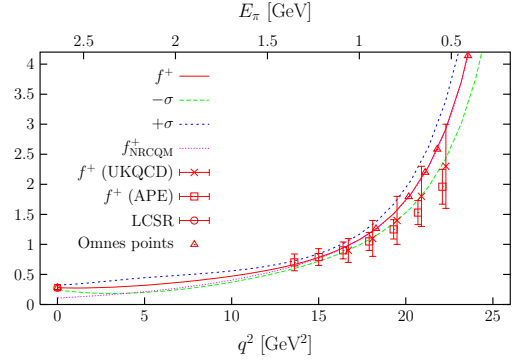}}
\caption{Omn\`es improved form factor (solid line). 
The subtraction points are denoted by
triangles. The $\pm\sigma$ lines show the theoretical uncertainty band. }
\label{fig2}
\end{center}
\end{figure}

Figure~\ref{fig2} shows with a solid line the form factor obtained using the Omn\`es
representation with six subtraction points: 
we take five $q^2$ values between $18$ GeV$^2$ and $q^2_{\rm max}$ for which we 
use the 
$f^+$ NRCQM predictions
(valence + $B^*$ pole),
plus the LCSR prediction at $q^2=0$. 
The $\pm\sigma$ lines enclose a $68\%
$ confidence level region that we have obtained from an estimation of the
theoretical uncertainties. The latter have two origins:
 (i) uncertainties in the quark--antiquark
nonrelativistic interaction and (ii) uncertainties on the product
$g_{B^*B\pi}f_{B^*}$, and on the input to the multiply subtracted
Omn\`es representation.  See
Ref.~\cite{1} for details. 

By Comparison with the experimental value for the decay
width, we obtain
\begin{equation}
|V_{ub}|=0.0034\pm 0.0003({\rm exp.}) \pm 0.0007({\rm theo.})
\end{equation}
in very good agreement with the value found by the CLEO Collaboration~\cite{4}.
\section{$D\to\pi l \bar{\nu}_l$ and $D\to K l \bar{\nu}_l$}
\begin{figure}
\begin{center}
\resizebox{0.75\columnwidth}{!}{\includegraphics{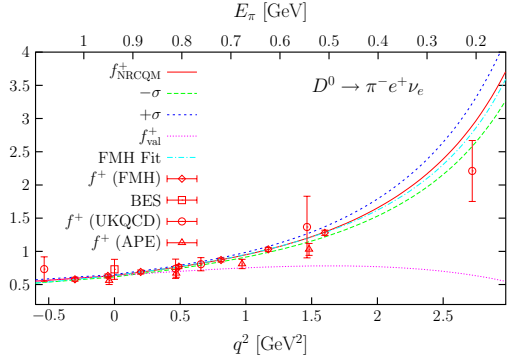}}
\caption{\label{fig3} The solid line denotes our determination of the
$f^+$ form factor ($f^+_{NRCQM}$) for the $D^0\to\pi^- e^+ \nu_e$
decay. The $\pm\sigma$ lines denote the theoretical uncertainty band
on the form factor. We compare with experimental data by the BES Collaboration
\cite{5} and with lattice results by the Fermilab-MILC-HPQCD~\cite{fermi},
UKQCD~\cite{ukqcd2} and APE ~\cite{ape} Collaborations.}
\end{center}
\end{figure}

\begin{figure}
\begin{center}
\resizebox{0.75\columnwidth}{!}{\includegraphics{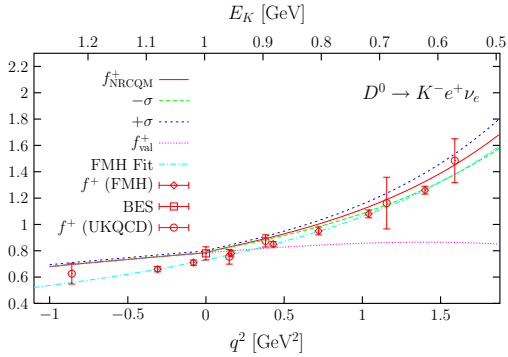}}
\caption{\label{fig4} Same as Fig.~\ref{fig3} for the decay $D^0\to
K^- e^+ \nu_e$}
\end{center}
\end{figure}

Our results for the $f^+$ form factor are depicted in
Figures~\ref{fig3} and \ref{fig4}. As before we have considered valence quark
plus resonant pole contributions ($D^*$ and $D_s^*$ respectively). In
both cases, we obtain a good description in the physical region of the experimental data~\cite{5} and 
previous lattice results~\cite{fermi,ukqcd2,ape}, without  using
 the Omn\`es dispersion relation. In the case of the 
 $D\to K$ decay, our predictions for  negative $q^2$  values could had been improved
by  the Omn\`es representation.  

\begin{figure}
\begin{center}
\resizebox{0.8\columnwidth}{!}{\includegraphics{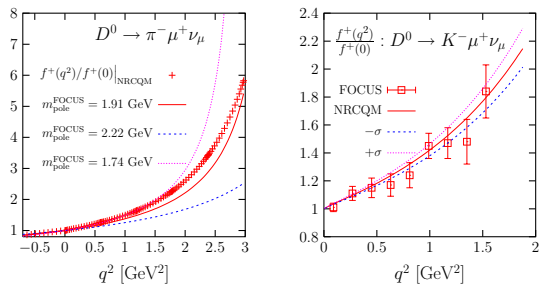}}
\caption{\label{fig5} NRCQM predictions for the ratio
$f^+(q^2)/f^+(0)$ for $D\to\pi$ and $D\to K$ decays. We compare with
experimental resuls by the FOCUS
Collaboration \cite{6}  (a pole fit
($m_{pole}=1.91^{+0.31}_{-0.17}$\,GeV) to  data in the $D\to\pi$ case). For the
$D\to K$ case we show the theoretical uncertainty band.}
\end{center}
\end{figure}

In
Fig.~\ref{fig5} we  compare our
results for the $f^+(q^2)/f^+(0)$ with experimental results by the FOCUS
Collaboration~\cite{6}.
 We find very good agreement with the data. 
 
  Besides we have found
 for the
decay widths 
\begin{eqnarray}
&\Gamma(D^0\to\pi^-e^+\nu_e)=(5.2\pm0.1({\rm exp.})\pm0.5 ({\rm theo.}))
\nonumber\\
&\hspace{3.25cm}\times 
10^{-12} {\rm MeV}\nonumber\\
&\Gamma(D^0\to K^-e^+\nu_e)=(66\pm 3 ({\rm theo.}))
\times 10^{-12} {\rm MeV}
\end{eqnarray}
For $D\to \pi$ we are in good agreement with experimental data while for 
$D\to K$ our result is two standard deviations higher.
\section{Concluding remarks}
We have shown the limitations of a pure valence quark model to describe the
$B\to\pi$, $D\to\pi$ and $D\to K$ semileptonic decays. As a first
correction, we have included vector resonance pole contributions which dominate
 the relevant $f^+$ form factor at high $q^2$ transfers.
 Subsequently, for the $B\to \pi$ decay, we have applied a multiply
subtracted Omn\`es dispersion relation. This has allowed us to extend the
results of the NRCQM model to the whole $q^2$ range.
 Our result for $|V_{ub}|$
 is in good agreement with recent experimental data by the CLEO
Collaboration. For $f^+(q^2)$ of the $D\to\pi$ and $D\to K$ decays and $q^2$ in
the physical region we have found good
agreement with experimental and lattice data. 

\section{Acknowledgments}
This work was supported by DGI and FEDER funds, under Contracts
No. FIS2005-00810, BFM2003-00856 and FPA2004-05616, by the
Junta de Andaluc\'\i a and Junta de Castilla y Le\'on under Contracts
No. FQM0225 and No. SA104/04, and it is a part of the EU integrated
infrastructure initiative Hadron Physics Project under Contract
No. RII3-CT-2004-506078. J.M.V.-V. acknowledges an E.P.I.F contract
with the University of Salamanca. C. A. acknowledges a research
contract with the University of Granada.

\end{document}